\documentclass[10pt, conference]{IEEEtran}
\IEEEoverridecommandlockouts
\usepackage{cite}
\usepackage{amsmath,amssymb,amsfonts}
\usepackage{graphicx}
 \usepackage{tikz}
\usepackage{subcaption}
\usepackage{textcomp}
\usepackage{xcolor}

\def\BibTeX{{\rm B\kern-.05em{\sc i\kern-.025em b}\kern-.08em
    T\kern-.1667em\lower.7ex\hbox{E}\kern-.125emX}}

\begin{document}
\title{Toward a Multi-Layer ML-Based Security Framework for Industrial IoT}

\author{
	\IEEEauthorblockN{Aymen Bouferroum}
	\IEEEauthorblockA{
		\textit{Inria Lille-Nord Europe}\\
		Lille, France\\
		aymen-salah-eddine.bouferroum@inria.fr
	}
	
	\and
	\IEEEauthorblockN{Valeria Loscri}
	\IEEEauthorblockA{
		\textit{Inria Lille-Nord Europe}\\
		Lille, France\\
		valeria.loscri@inria.fr
	}
	
	\and
	\IEEEauthorblockN{Abderrahim Benslimane}
	\IEEEauthorblockA{
		\textit{LIA/CERI University of Avignon}\\
		Avignon, France\\
		abderrahim.benslimane@univ-avignon.fr
	}
}

\maketitle

\begin{abstract}
The Industrial Internet of Things (IIoT) introduces significant security challenges as resource-constrained devices become increasingly integrated into critical industrial processes. Existing security approaches typically address threats at a single network layer, often relying on expensive hardware and remaining confined to simulation environments. In this paper, we present the research framework and contributions of our doctoral thesis, which aims to develop a lightweight, Machine Learning (ML)-based security framework for IIoT environments. We first describe our adoption of the Tm-IIoT trust model and the Hybrid IIoT (H-IIoT) architecture as foundational baselines, then introduce the Trust Convergence Acceleration (TCA) approach, our primary contribution that integrates ML to predict and mitigate the impact of degraded network conditions on trust convergence, achieving up to a 28.6\% reduction in convergence time while maintaining robustness against adversarial behaviors. We then propose a real-world deployment architecture based on affordable, open-source hardware, designed to implement and extend the security framework. Finally, we outline our ongoing research toward multi-layer attack detection, including physical-layer threat identification and considerations for robustness against adversarial ML attacks.
\end{abstract}


\section{Introduction}

The rapid advancement of the Industrial Internet of Things (IIoT) through Industry 4.0 initiatives \cite{sadeghi2015security} has led to the widespread integration of cyber-physical systems, cloud computing, and artificial intelligence into industrial processes, enhancing operational efficiency and decision-making capabilities \cite{boudagdigue2020trust}. This evolution, however, introduces significant security challenges, particularly in environments where resource-constrained devices must operate reliably under harsh industrial conditions \cite{serror2021challenges}. The heterogeneous nature of IIoT devices, with their limited computational and energy resources, demands security solutions that remain lightweight while providing effective protection across diverse deployment scenarios.

IIoT systems are exposed to a wide spectrum of threats spanning multiple layers of the network stack \cite{tsiknas2021cyber, panchal2018security}. At the application layer, compromised or malicious nodes may provide inaccurate data, refuse to cooperate in network activities, or conduct reputation attacks to undermine legitimate devices. At the network and physical layers, adversaries may disrupt wireless communication channels through jamming, routing manipulation, or signal interference \cite{tsiknas2021cyber}. These challenges are compounded by the harsh electromagnetic environment inherent to industrial facilities, where emissions from heavy machinery, electric motors, and welding equipment, as well as multipath propagation caused by metallic structures and dense equipment layouts, severely degrade wireless signal quality \cite{wetzker2016troubleshooting}. Despite this multi-layer threat landscape, the majority of existing security solutions address a single layer in isolation, often depend on expensive proprietary hardware, and remain confined to simulation-based evaluations without consideration for real-world deployment on constrained devices.

Trust management has emerged as a compelling mechanism for securing IIoT environments, as it enables continuous assessment of device behavior without the computational overhead associated with complex cryptographic operations \cite{boudagdigue2018distributed}. While trust-based decisions are typically enforced at the application layer, the underlying trust evaluation inherently relies on cross-layer observations: cooperation metrics monitor packet forwarding behavior at the network layer, and Quality of Service (QoS) indicators such as Signal-to-Noise Ratio (SNR), latency, and jitter capture conditions at the physical and transport layers. However, existing trust models typically assume stable network conditions, overlooking the significant impact that dynamic network quality fluctuations have on trust evaluation accuracy and convergence speed \cite{Mohammadi2019}. This limitation creates extended periods of uncertainty during which the system cannot reliably distinguish between legitimate and malicious nodes, leaving the network vulnerable.

In this paper, we present the research framework of our doctoral thesis, conducted within the scope of the PEPR Future Networks program under the France 2030 initiative, and specifically as part of the FITNESS project \cite{cassiau2025fitness}, which aims to overcome the technical hurdles of IoT adoption through a collaborative, multi-partner research effort. Our thesis focuses on developing a lightweight, ML-based, multi-layer security framework for IIoT. The research began by selecting the Tm-IIoT trust model \cite{boudagdigue2020trust} and its Hybrid IIoT (H-IIoT) architecture as foundational baselines, then progressively extended this foundation to address network degradation, real-world deployment, and multi-layer threat coverage. Our work addresses the following objectives:

\begin{itemize}
    \item Design of the Trust Convergence Acceleration (TCA) approach \cite{bouferroum2025tca}, an adaptive ML-enhanced trust management solution that accelerates trust convergence under degraded network conditions.
    \item Proposal of a real-world deployment architecture using cost-effective, open-source hardware that concretely implements the H-IIoT trust model and enables practical validation.
    \item Progressive extension of the framework toward multi-layer attack detection and response, including physical-layer threat identification and robustness against adversarial ML attacks.
\end{itemize}

The remainder of this paper is organized as follows: Section~II introduces the context and positioning of our research. Section~III details the TCA approach and its performance evaluation. Section~IV presents our proposed deployment architecture and outlines ongoing research directions. Section~V concludes the paper.

\section{Context and Positioning}

\subsection{IIoT Network Architecture and Trust Baseline}

\begin{figure}[t]
\centering
\includegraphics[scale=0.19]{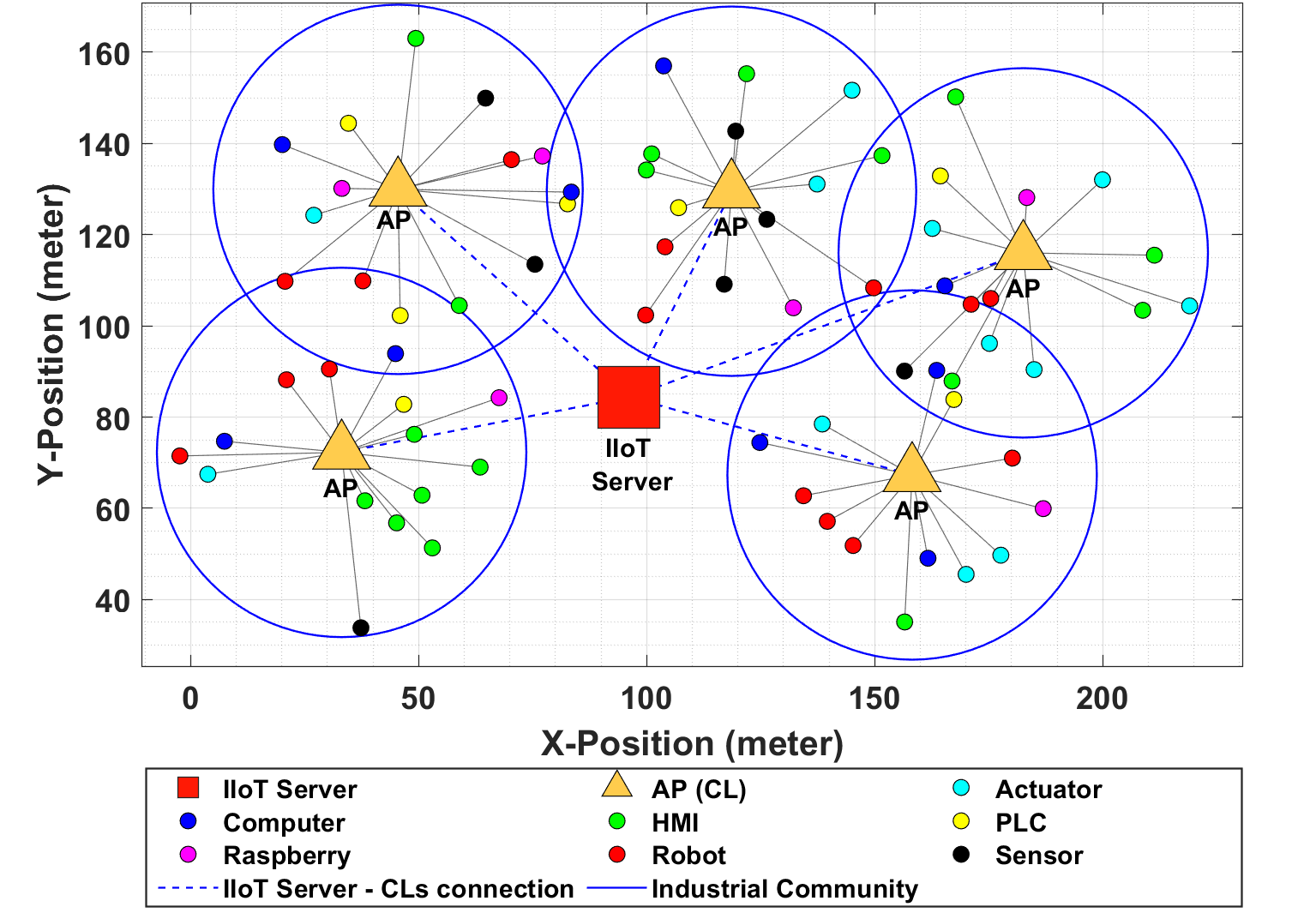}
\caption{H-IIoT network architecture.}
\label{fig:network_architecture}
\vspace{-10pt}
\end{figure}

Our research builds upon the H-IIoT architecture \cite{boudagdigue2020trust}, which organizes IIoT devices into managed industrial communities to balance localized trust management with global network oversight. The architecture comprises three hierarchical components: an IIoT Server acting as a cloud-based central authority for global trust aggregation, Community Leaders (CLs) serving as trusted supervisors for their respective communities, and Member Nodes (MNs) representing the diverse IIoT devices (sensors, actuators, Programmable Logic Controllers (PLCs), collaborative robots) that form each community. Fig.~\ref{fig:network_architecture} illustrates this hierarchical organization.

Within this architecture, the Tm-IIoT trust model \cite{boudagdigue2020trust} evaluates each MN's trustworthiness through a trust metric ($Tm$) that aggregates three behavioral indicators: the cooperation rate ($C_{MN_i}$), which measures packet forwarding reliability; direct honesty ($D_{MN_i}$), which assesses adherence to expected operational roles; and indirect honesty ($I_{MN_i}$), which captures peer-provided reputation feedback. Trust evolution is modeled as a finite-state discrete-time Markov chain with 11 states (trust values from 0 to 1 in increments of 0.1), enabling computationally efficient evaluation suitable for resource-constrained IIoT devices. We selected Tm-IIoT as our baseline for its demonstrated accuracy in detecting malicious nodes, its low computational footprint, and the efficiency of the H-IIoT hierarchy in distributing the trust management burden across CLs while maintaining network-wide consistency through the central server.

Our implementation leverages Wi-Fi 6 (IEEE 802.11ax) \cite{wba2022wifi, IEEE80211ax}, whose features (OFDMA, MU-MIMO, TWT) are well suited for dense, energy-constrained industrial environments.

\subsection{Trust Management in IIoT}

Trust management provides a mechanism for assessing device reliability in IIoT networks, continuously monitoring node behavior and assigning trust levels based on observed interactions \cite{boudagdigue2018distributed, grandison2000survey}. Trust management approaches can be broadly categorized into centralized, distributed, and hybrid models \cite{alshehri2018clustering, chen2015trust}. Centralized approaches offer global visibility but face scalability limitations. Distributed models improve scalability at the cost of increased convergence time and energy consumption \cite{abderrahim2017tmcoi}. Hybrid models, such as H-IIoT, combine the advantages of both by enabling localized management within communities while preserving a global perspective through the central server.

A critical limitation shared by existing approaches is the lack of consideration for real-time network conditions. In operational IIoT environments, network quality is subject to continuous fluctuations caused by electromagnetic interference from heavy machinery, electric motors, and welding equipment, as well as multipath propagation from metallic structures and dense equipment layouts \cite{wetzker2016troubleshooting}. While some researchers have employed recommendation filtering algorithms \cite{CHEN2021107952} or introduced fixed probability models \cite{boudagdigue2020trust} to address contextual variability, these approaches fail to capture the dynamic relationship between network quality and trust convergence speed, leaving the system vulnerable during periods of degraded connectivity.

Moreover, trust management alone does not address the full IIoT threat landscape, as these systems face attacks at multiple layers of the network stack \cite{tsiknas2021cyber}. Physical-layer attacks such as jamming and signal interference can disrupt wireless communications \cite{hassan2023realtime}, while network-layer attacks may exploit routing protocols or communication channels \cite{panchal2018security}. A comprehensive security framework for IIoT must therefore progressively address threats across these layers rather than treating them in isolation. Our thesis adopts this multi-layer perspective, beginning with trust management as a foundational component that inherently draws on cross-layer observations, and designing the architecture from the outset for extensibility toward broader threat coverage.

\section{Trust Convergence Acceleration}

The TCA approach \cite{bouferroum2025tca} constitutes our primary contribution, addressing the challenge of maintaining effective trust management under the dynamic network conditions inherent to industrial environments. The complete algorithmic details, dataset generation methodology, and extended evaluation are available in \cite{bouferroum2025tca}; we summarize the key elements below.

\subsection{Problem Formulation}

To quantify the impact of network quality on trust evaluation, we define a composite Network Condition parameter ($netC$) that aggregates key QoS metrics identified as significant for network performance \cite{Sugeng2015THEIO}:
    \vspace{-5pt}

\begin{equation}
    \label{eq:netC}
    \scalebox{0.88}{$\begin{gathered} netC = \alpha \cdot SNR_{norm} + \beta \cdot (1 - PL_{norm}) + \gamma \cdot (1 - J_{norm}) \\ + \delta \cdot (1 - L_{norm}) + \tau \cdot T_{norm} + \sigma \cdot SINR_{norm}, \end{gathered}$}
\end{equation}

where the normalized metrics capture SNR, Packet Loss (PL), Jitter (J), Latency (L), Throughput (T), and Signal-to-Interference-plus-Noise Ratio (SINR). For metrics where lower values indicate better performance, complements are used to ensure consistent polarity, and equal weights ($1/6$ each) are assigned to reflect the balanced contribution of each parameter.

Simulation analysis using MATLAB's WLAN and Communications Toolboxes with realistic IEEE 802.11ax channel conditions \cite{bouferroum2025tca} reveals that poor network conditions extend trust convergence from 4 to 12 time units compared to good conditions, creating prolonged vulnerability windows where malicious nodes remain undetected.

\subsection{ML-Based Prediction and Dynamic Adjustment}

\begin{figure}[t]
    \centering
    \includegraphics[width=0.97\linewidth]{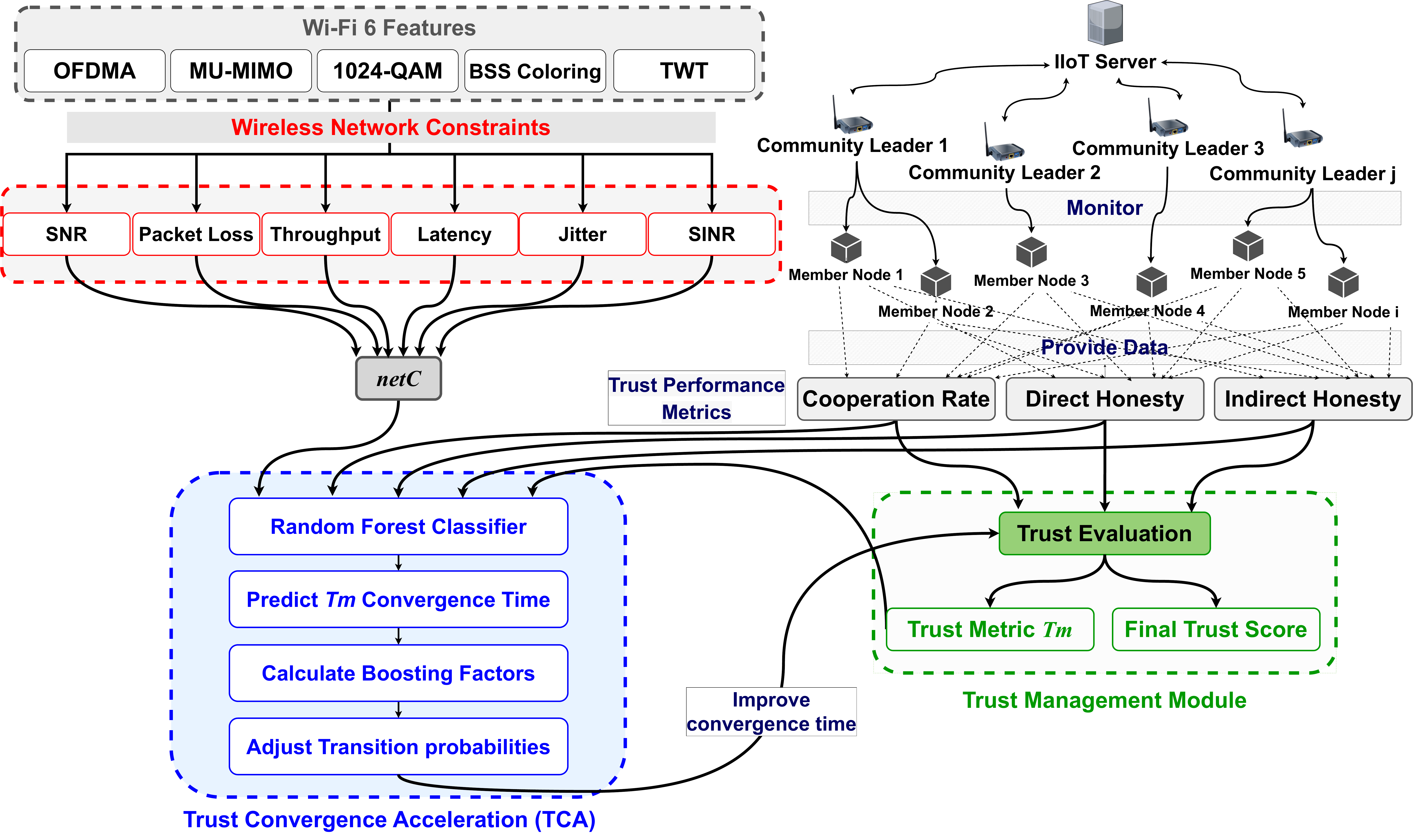}
    \caption{Architecture of the TCA solution \cite{bouferroum2025tca}.}
    \label{fig:tca_solution}
    \vspace{-10pt}
\end{figure}

The TCA solution enhances trust management adaptability through three integrated components, as illustrated in Fig.~\ref{fig:tca_solution}: network condition quantification via $netC$, ML-based convergence time prediction, and dynamic trust adjustment through a boosting factor.

A Random Forest classifier \cite{Breiman2001} predicts the expected trust convergence time based on network conditions and trust parameters. Selected after comparative evaluation against three alternative models (details in \cite{bouferroum2025tca}), it achieves 92.66\% accuracy and 92.63\% F1-score \cite{chicco2020advantages}, with robustness to noise and low inference overhead suitable for resource-constrained platforms.

Based on the ML predictions, TCA dynamically adjusts trust convergence through a boosting factor ($bf$) that modifies transition probabilities in the underlying Markov chain trust model \cite{boudagdigue2020trust}:

\vspace{-10pt}
\begin{equation}
    \scalebox{0.9}{$
    \begin{gathered}
        bf = 1 + (1 - netC) \cdot \lambda \cdot \min\left(\frac{pc}{maxT}, 1\right),
    \end{gathered}$}
\end{equation}

where $pc$ is the predicted convergence class, $maxT$ the maximum convergence time, and $\lambda$ an empirically tuned scaling factor (0.2). When $netC \geq 0.8$, no boosting is applied ($bf = 1$), ensuring selective intervention only under degraded conditions. This design distinguishes network-induced degradation from genuinely malicious behavior, since boosting is driven by network metrics rather than trust metrics.

\subsection{Performance Evaluation}

\begin{figure}[t]
    \centering
    \includegraphics[width=1\linewidth]{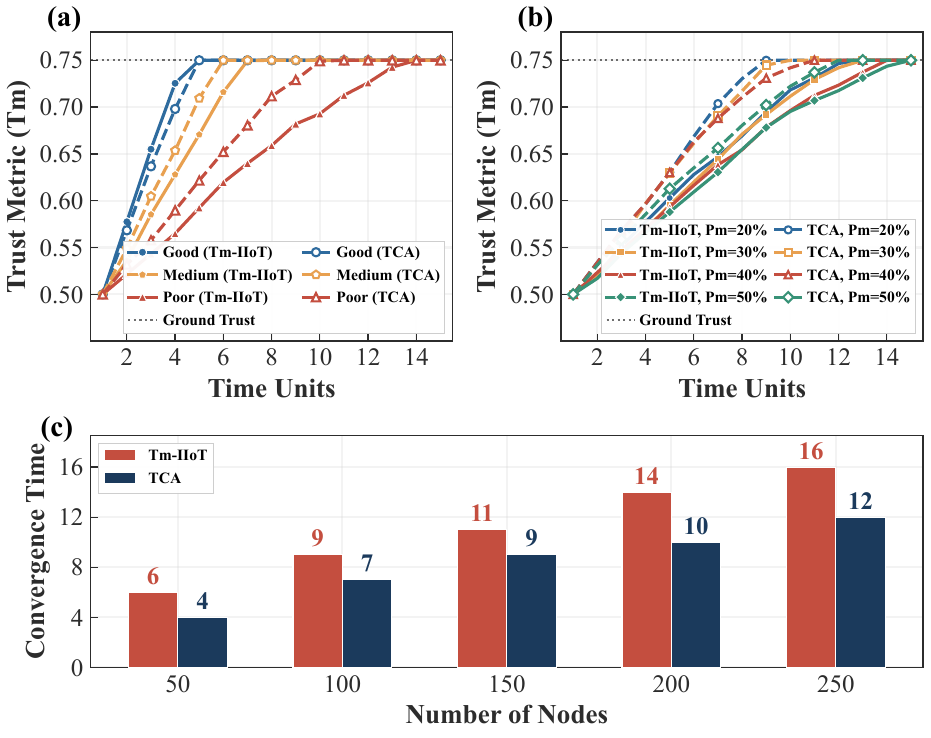}
    \caption{TCA performance evaluation: (a) trust convergence under Good, Medium, and Poor network conditions comparing Tm-IIoT and TCA, (b) convergence resilience under varying malicious node ratios ($P_m$ = 20\%--50\%), and (c) scalability analysis of average convergence time with increasing network size.}
    \label{fig:tca_results}
    \vspace{-15pt}
\end{figure}

Fig.~\ref{fig:tca_results} presents the key results against the baseline Tm-IIoT
model. In terms of convergence speed (Fig.~\ref{fig:tca_results}a), TCA achieves
up to 28.6\% reduction under poor conditions (10 vs 14 time units), 14.3\% under
medium conditions, and equivalent performance under good conditions, confirming
selective activation. Under bad-mouthing attacks~\cite{buja2023enhancing}
(Fig.~\ref{fig:tca_results}b), TCA reduces convergence time by 30.77\% at
$P_m$=20\%, with improvements persisting up to $P_m$=50\% (20\%) while preserving
final trust accuracy. Fig.~\ref{fig:tca_results}c confirms 18\% to 33\% faster
convergence across 50 to 250 nodes, validating practical scalability.

\section{Proposed Deployment and Research Directions}

\subsection{Real-World Deployment Architecture}

\begin{figure}[t]
    \centering
    \includegraphics[width=1\linewidth]{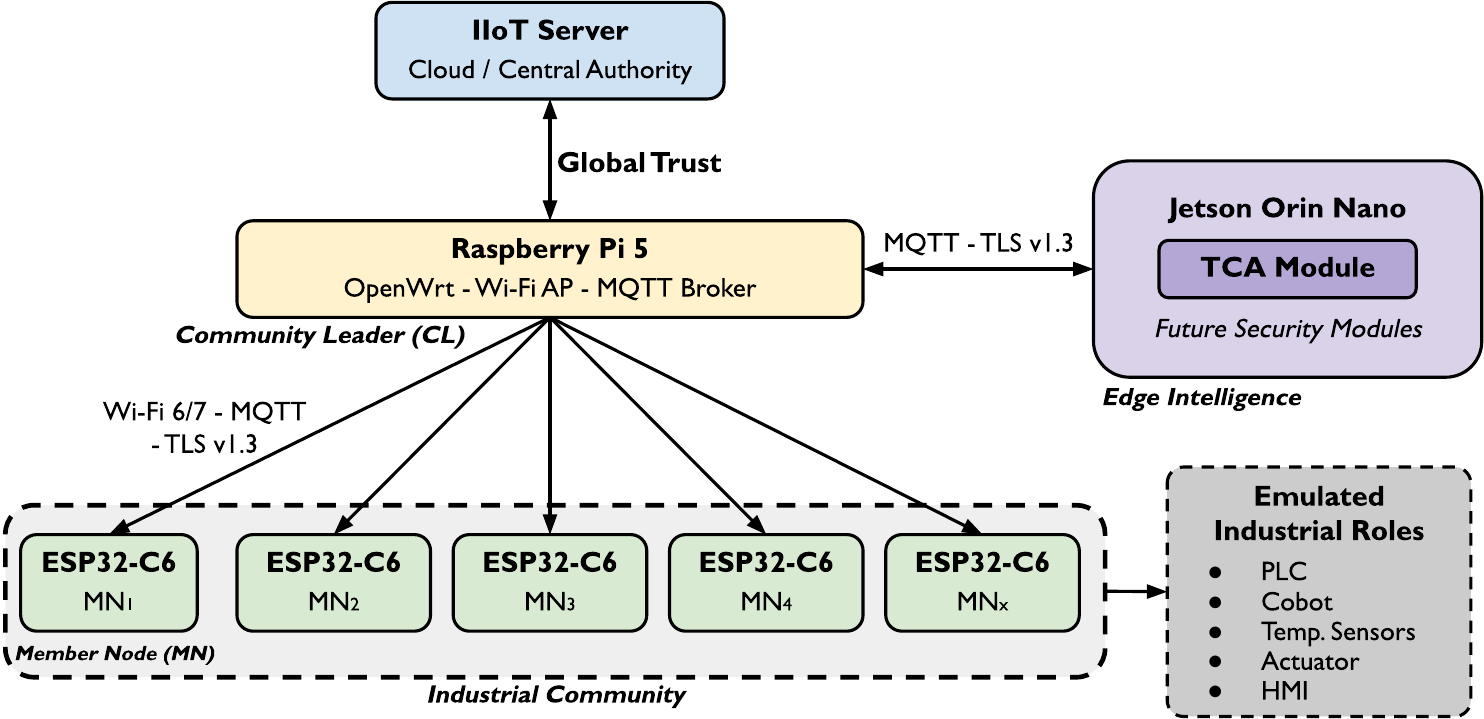}
    \caption{Proposed real-world IIoT deployment architecture.}
    \label{fig:testbed_architecture}
    \vspace{-10pt}
\end{figure}

To transition our framework from simulation to practical deployment, we propose a concrete IIoT testbed architecture that maps the H-IIoT model onto affordable, commercially available hardware, as illustrated in Fig.~\ref{fig:testbed_architecture}.

The proposed architecture faithfully reproduces the H-IIoT hierarchy using three operational layers. At the MN level, Wi-Fi 6 capable microcontrollers (ESP32-C6) serve as IIoT device representatives, emulating the communication behavior and data reporting patterns of industrial equipment such as PLCs, collaborative robots, environmental sensors, and actuators. Equipped with diverse sensing capabilities (e.g., temperature, humidity...), these microcontrollers generate realistic industrial data streams while preserving the resource constraints and wireless characteristics of real deployments. This proxy-based approach is a well-established practice in IIoT security research, enabling controlled and reproducible experimentation. At the CL level, a Raspberry Pi 5 running OpenWrt serves as the Wi-Fi access point and MQTT message broker, managing communication and coordination within each industrial community. At the edge intelligence level, an NVIDIA Jetson Orin Nano provides GPU-accelerated ML inference through its NVIDIA Ampere architecture (1024 CUDA cores, 32 Tensor Cores), delivering up to 67 TOPS of AI performance within a configurable power envelope of 7 to 25~W. This platform hosts the pre-trained TCA module for real-time trust convergence optimization and serves as the designated platform for deploying additional security components in ongoing research, including computationally intensive deep learning models for multi-layer attack classification.
Communication across the architecture relies on the MQTT protocol \cite{sethi2017internet} over Wi-Fi 6, secured with TLS v1.3 for data confidentiality and integrity. The entire software stack is built on open-source components, and the total hardware cost for a representative community remains around \texteuro800, making the architecture fully reproducible and accessible for the research community.

\subsection{Malicious Node Response and Remediation}

An effective security framework must not only evaluate node trustworthiness but also implement appropriate response mechanisms when malicious behavior is detected. Within the H-IIoT hierarchy enhanced by TCA, when a MN's trust metric falls below the defined threshold, the CL isolates the suspected node from critical network activities, preventing it from influencing cooperative tasks and the trust evaluations of other community members. Simultaneously, an alert is propagated to the IIoT Server to update the global trust view and coordinate cross-community awareness. To avoid permanent exclusion of devices experiencing transient faults rather than genuinely malicious intent, a quarantine-and-reevaluation procedure is employed: the isolated node undergoes a monitored observation period and, if subsequent evaluations confirm legitimate behavior, it is progressively reintegrated with appropriate monitoring. This graduated response mechanism ensures proportional and reversible actions, which is particularly important in industrial settings where device availability directly impacts operational continuity.

\subsection{Extension to Multi-Layer Attack Detection}

Our ongoing research extends the security framework beyond trust management toward multi-layer attack detection. The modular architecture of the proposed deployment is specifically designed to accommodate additional security components targeting threats at the network and physical layers \cite{tsiknas2021cyber, panchal2018security}. At the physical layer, we are actively investigating ML-based techniques for detecting jamming and interference attacks by exploiting wireless channel characteristics available on the deployed hardware \cite{hassan2023realtime}. The GPU-accelerated inference capabilities of the Jetson Orin Nano enable the deployment of these detection models alongside the existing TCA module. A key consideration in this extension is robustness against adversarial ML attacks; as the framework increasingly relies on ML models for security decisions, ensuring that these models resist adversarial manipulation (e.g., evasion attacks, poisoning of training data) becomes essential.
This progressive approach allows each security component to operate independently while sharing information through the centralized trust framework. Observations from lower network layers can inform and refine the trust evaluation process, enabling the system to distinguish between genuine device misbehavior and performance degradation caused by external attacks. By maintaining the same low-cost hardware infrastructure, we ensure that the multi-layer extension preserves the accessible and replicable nature of the overall framework.
\vspace{-5pt}

\section{Conclusion}

In this paper, we have presented the research framework of our doctoral thesis, which addresses the need for a lightweight, ML-based, multi-layer security approach tailored to IIoT environments. Building upon the Tm-IIoT trust model and the H-IIoT architecture, we developed the TCA approach \cite{bouferroum2025tca}, which dynamically adapts trust evaluation to network condition fluctuations, achieving up to 28.6\% faster convergence with demonstrated resilience against adversarial behaviors across varying attack intensities and network scales. Although our current implementation leverages IEEE 802.11ax (Wi-Fi 6), the framework relies on technology-agnostic QoS metrics (SNR, latency, jitter, throughput), enabling its application across diverse network architectures including emerging 5G and 6G industrial infrastructures, and interfacing with standard protocols such as TCP/IP or UDP. The proposed deployment architecture maps this framework onto affordable, open-source hardware, providing a replicable foundation for real-world IIoT security research. Our ongoing work targets validation with real-world data, multi-layer attack detection with robustness against adversarial ML attacks, and progressive extension toward end-to-end security for industrial IoT systems.

\section*{Acknowledgment}
This work was funded by the French National Research Agency (ANR-22-PEFT-0007) as part of France 2030 and the NF-FITNESS project \cite{cassiau2025fitness}.

\bibliographystyle{ieeetr}
\bibliography{bibfile}
\end{document}